\documentclass[11pt]{book}

\usepackage{amsmath,amsthm,amscd,amssymb}
\usepackage{latexsym}

\theoremstyle{definition}
\newtheorem*{abstract}{Abstract}

\begin{document}

\chapter{Gravitoelectromagnetism:\\ A Brief Review}

Bahram Mashhoon\\Department of Physics and
Astronomy\\University of Missouri-Columbia\\Columbia,
Missouri 65211, USA

\begin{abstract} The main theoretical aspects of 
gravitoelectromagnetism (``GEM'') are presented. Two basic approaches 
to this
subject are described and the role of the gravitational Larmor 
theorem is emphasized. Some of the consequences of GEM are
briefly mentioned.\end{abstract}

\section{Introduction}

The analogy between gravitation and electromagnetism has a long 
history. The similarity between Newton's law of gravitation
and Coulomb's law of electricity naturally led to a gravitoelectric 
description of Newtonian gravitation. Moreover,
on the basis of advances in electrodynamics in the second half of the 
nineteenth century, Holzm\"uller \cite{1} and Tisserand
\cite{2} postulated that the gravitational force exerted by the Sun 
on the planets of the solar system had an additional
``magnetic'' component. This extra force led to the precession of the 
planetary orbits; therefore, it could be adjusted in
order to account for the excess perihelion precession of Mercury. 
Decades later, however, Einstein's general relativity
provided a beautiful explanation of the excess motion of Mercury's 
perihelion in terms of a relativistic gravitoelectric
correction to the Newtonian gravitational potential of the Sun 
\cite{3}. Furthermore, general relativity, which is a field
theory of gravitation, contains a gravitomagnetic field due to mass 
current \cite{4}. Indeed, to bring together Newtonian
gravitation and Lorentz invariance in a consistent field-theoretic 
framework, the introduction of a gravitomagnetic field is
unavoidable.

According to general relativity, the proper rotation of the Sun 
produces a gravitomagnetic field and the influence of this
field on planetary orbits was first considered by de Sitter \cite{5} 
and later in a more general form by Lense and Thirring
\cite{4}. The gravitomagnetic contribution to the excess motion of 
Mercury's perihelion turns out to be much smaller and in
the opposite sense compared to the main gravitoelectric motion; in 
fact, it turns out that the Lense-Thirring precession of
planetary orbits is too small to be measurable at present. On the 
other hand, evidence for the gravitomagnetic field of the
Earth has been offered by Ciufolini by studying the motion of 
laser-ranged satellites LAGEOS and LAGEOS II \cite{6}.
The precise measurement of this field via superconducting gyroscopes 
in a drag-free satellite in polar orbit about the Earth
is one of the aims of NASA's GP-B \cite{7}.

Within the framework of general relativity, gravitoelectromagnetism 
(``GEM'') has been discussed by a number of authors
\cite{3,8}; a more extensive list of references is provided in 
\cite{9}. The purpose of this review is to present the two
principal approaches to GEM and briefly describe some of their consequences.

\section{Linear Perturbation Approach to GEM}

We are interested in the general linear solution of the gravitational 
field equations \cite{3}. It is assumed that a global
background inertial frame with coordinates $x^\mu =(ct,{\bf x})$ and 
Minkowski metric $\eta_{\mu \nu}$ is perturbed due to the
presence of gravitating sources such that 
$g_{\mu\nu}=\eta_{\mu\nu}+h_{\mu\nu}(x)$. It proves useful to define 
the
trace-reversed amplitude 
$\bar{h}_{\mu\nu}=h_{\mu\nu}-\frac{1}{2}\eta_{\mu\nu}h$, where 
$h=\eta^{\mu\nu}h_{\mu\nu}$ is the
trace of $h_{\mu\nu}$. To linear order in the perturbation, 
Einstein's field equations
\begin{equation}\label{eq1} 
R_{\mu\nu}-\frac{1}{2}g_{\mu\nu}R=\frac{8\pi 
G}{c^4}T_{\mu\nu}\end{equation}
take the form
\begin{equation}\label{eq2} \Box \bar{h}_{\mu\nu}=-\frac{16\pi 
G}{c^4}T_{\mu\nu}\end{equation}
after imposing the transverse gauge condition 
$\bar{h}^{\mu\nu}_{\;\;\;\; ,\nu}=0$. The general solution of 
\eqref{eq2} is a
superposition of a particular solution together with the general 
solution of the wave equation; however, we are only
interested in the special retarded solution of \eqref{eq2} given by
\begin{equation}\label{eq3} 
\bar{h}_{\mu\nu}=\frac{4G}{c^4}\int\frac{T_{\mu\nu}(ct-|{\bf
x}-{\bf x}'|,{\bf x}')}{|{\bf x}-{\bf x}'|}d^3x'.\end{equation} Let 
us define the matter density $\rho$ and matter current
${\bf j}=\rho {\bf v}$ via
$T^{00}=\rho c^2$ and $T^{0i}=cj^i$, respectively. Moreover, it is 
useful to define the GEM potentials $\Phi$ and ${\bf A}$ in
terms of $\rho$ and ${\bf j}$ as
$\bar{h}_{00}=4\Phi /c^2$ and $\bar{h}_{0i}=-2A_i/c^2$, respectively. 
Assuming that the source consists of a finite
distribution of slowly moving matter with $|{\bf v}|<<c$, $T_{ij}\sim 
\rho v_iv_j+p\delta_{ij}$, where $p$ is the pressure, and
\eqref{eq3} imply that
$\bar{h}_{ij}=O(c^{-4})$. All terms of $O(c^{-4})$ will be neglected 
in this analysis. Under these conditions, the spacetime
metric has the GEM form
\begin{equation}\label{eq4} 
ds^2=-c^2\left(1-2\frac{\Phi}{c^2}\right)dt^2-\frac{4}{c}({\bf 
A}\cdot d{\bf
x})dt+\left(1+2\frac{\Phi}{c^2}\right) \delta_{ij}dx^idx^j.\end{equation}
In the Newtonian limit, $\Phi$ reduces to the Newtonian gravitational 
potential, while ${\bf A}=O(c^{-1})$. If the source
distribution is confined around the origin of spatial coordinates, 
then far from the source
\begin{equation}\label{eq5} \Phi \sim \frac{GM}{r},\quad {\bf A}\sim 
\frac{G}{c}\frac{{\bf J}\times {\bf
x}}{r^3},\end{equation}
where $r=|{\bf x}|$ and $M$ and ${\bf J}$ are the mass and angular 
momentum of the source, respectively. Moreover, the
transverse gauge condition reduces to
\begin{equation}\label{eq6} \frac{1}{c}\frac{\partial \Phi}{\partial 
t}+\nabla \cdot \left(\frac{1}{2}{\bf
A}\right)=0.\end{equation}

We define the GEM fields via
\begin{equation}\label{eq7} {\bf E}=-\nabla \Phi 
-\frac{1}{c}\frac{\partial}{\partial t}\left( \frac{1}{2}{\bf 
A}\right),
\quad {\bf B}=\nabla \times {\bf A},\end{equation}
in direct analogy with electromagnetism. It follows from these 
definitions that the GEM fields have dimensions of acceleration
and
\begin{equation}\label{eq8} \nabla \times {\bf 
E}=-\frac{1}{c}\frac{\partial }{\partial t}\left( \frac{1}{2}{\bf
B}\right),\quad \nabla \cdot \left( \frac{1}{2}{\bf B}\right)=0.\end{equation}
Furthermore, \eqref{eq2} implies that
\begin{equation}\label{eq9} \nabla \cdot {\bf E}=4\pi G\rho ,\quad 
\nabla \times \left(\frac{1}{2}{\bf
B}\right)=\frac{1}{c}\frac{\partial }{\partial t} {\bf E}+\frac{4\pi 
G}{c}{\bf j}.\end{equation}
The GEM field equations \eqref{eq8} and \eqref{eq9} contain the 
continuity equation $\boldsymbol{\nabla}\cdot {\bf j}+\partial
\rho /\partial t=0$, as expected.

For a complete GEM theory, we need an analogue of the Lorentz force 
law. The Lagrangian for the motion of a test particle  of
mass $m$, $L=-mc ds/dt$, can be written to linear order in $\Phi$ and 
${\bf A}$ as
\begin{equation}\label{eq10} 
L=-mc^2\left(1-\frac{v^2}{c^2}\right)^{\frac{1}{2}}+m\gamma 
\left(1+\frac{v^2}{c^2}\right) \Phi
-\frac{2m}{c}\gamma {\bf v}\cdot {\bf A},\end{equation}
where $\gamma$ is the Lorentz factor. The equation of motion, $d{\bf 
p}/dt={\bf F}$, where ${\bf p}=\gamma m{\bf v}$ is the
kinetic momentum, takes a simple familiar form if $\partial {\bf 
A}/\partial t=0$ and ${\bf F}$ is expressed to lowest order
in ${\bf v}/c$, $\Phi$ and ${\bf A}$; then,
\begin{equation}\label{eq11}{\bf F}=-m{\bf E}-2m\frac{{\bf 
v}}{c}\times {\bf B}.\end{equation}
The canonical momentum of the particle is given in this case by ${\bf 
p}+(-2m/c){\bf A}$.

Let us now discuss the gauge freedom of the GEM potentials. Under a 
coordinate transformation $x^\mu\to
x^{'\mu}=x^\mu-\epsilon^\mu$, $h_{\mu\nu}\to 
h'_{\mu\nu}=h_{\mu\nu}+\epsilon_{\mu ,\nu}+\epsilon_{\nu, \mu}$ to 
linear order
in $\epsilon_\mu$. Therefore,
\begin{equation}\label{eq12} \bar{h}'_{\mu\nu}=\bar{h}_{\mu\nu}+\epsilon
_{\mu,\nu}+\epsilon_{\nu,\mu}-\eta_{\mu\nu}\epsilon^\alpha_{\;\; 
,\alpha}.\end{equation}
Under this transformation, the Riemann tensor remains invariant, but 
the connection changes. We must restrict $\epsilon_\mu$
in such a way that those elements of the connection defining GEM 
fields also remain invariant. The new metric perturbation
satisfies the transverse gauge condition as well, provided $\Box 
\epsilon _\mu=0$. Therefore, let $\epsilon_0=O(c^{-3})$ and
$\epsilon_i=O(c^{-4})$; then, \eqref{eq12} implies that
\begin{equation}\label{eq13} \Phi '=\Phi -\frac{1}{c}\frac{\partial 
}{\partial t}\Psi,\quad \frac{1}{2}{\bf A}'=\frac{1}{2}{\bf
A}+\nabla
\Psi ,\end{equation}
where $\Psi =c^2\epsilon ^0/4$ and $\Box \Psi=0$. Under the GEM gauge 
transformation \eqref{eq13}, the GEM fields are
invariant in close analogy with electrodynamics.

It is important to discuss the stress-energy tensor for GEM in the 
context of our approximation scheme. For the sake of
simplicity we set $c=G=1$ in what follows, unless specified otherwise. 
The Landau-Lifshitz pseudotensor $t_{\mu\nu}$ can be
employed to determine the local stress-energy content of the GEM 
fields. The general gauge-dependent result is somewhat
complicated and is given in \cite{10}; however, for a stationary 
configuration (i.e. $\partial \Phi /\partial t=0$ and
$\partial {\bf A}/\partial t=0$), one finds
\begin{align}\label{eq14} 4\pi 
Gt_{00}&=-\frac{7}{2}E^2+\sum_{i,j}A_{(i,j)} A_{(i,j)},\\
\label{eq15} 4\pi Gt_{0i}&=2({\bf E}\times {\bf B})_i,\\
\label{eq16} 4\pi 
Gt_{ij}&=\left(E_iE_j-\frac{1}{2}\delta_{ij}E^2\right)+\left( 
B_iB_j+\frac{1}{2}\delta
_{ij}B^2\right).\end{align}
There is some similarity between these and the corresponding 
relations in classical electrodynamics. In particular, the GEM
Poynting vector is given by
\begin{equation}\label{eq17} \mathcal{S}=-\frac{c}{2\pi G}{\bf 
E}\times {\bf B}.\end{equation}
For instance, gravitational energy circulates around a stationary 
source of mass $m$ and angular momentum ${\bf J}=J\hat{{\bf
z}}$ with a flow velocity
\begin{equation}\label{eq18} {\bf v}_g=k\frac{J}{Mr}\sin \theta 
\text{\boldmath$\hat{\phi}$\unboldmath}\end{equation}
in the same sense as the rotation of the mass. Here we employ 
spherical polar coordinates and $k=4/7$. The flow given by
\eqref{eq18} is divergence-free and the corresponding circulation is 
independent of the radial distance $r$ and is given by
$2\pi k(J/M)\sin ^2\theta$.

\section{Gravitational Larmor Theorem}

There is a certain degree of arbitrariness in the definitions of the 
GEM potentials and fields that leads to extraneous
numerical factors as compared to electrodynamics. To develop GEM in a 
way that would provide the closest possible connection
with the standard formulas of electrodynamics, we have adopted a 
convention that, among other things, provides a clear
statement of the gravitational Larmor theorem. We assume that a test 
particle of inertial mass $m$ has gravitoelectric charge
$q_E=-m$ and gravitomagnetic charge $q_B=-2m$. For a source that is a 
rotating body of mass $M$, the corresponding
charges are positive, $Q_E=M$ and $Q_B=2M$, respectively, in order to 
preserve the attractive nature of gravity. Moreover, the
ratio of the gravitomagnetic charge to the gravitoelectric charge is 
always $2$, since linearized gravity is a spin-$2$ field.
This is consistent with the fact that for a spin-$1$ field such as in 
Maxwell's theory, the corresponding ratio is always
unity.

The Larmor theorem originally established a basic local equivalence 
between magnetism and rotation \cite{11}. In fact, the
electromagnetic force on a test particle of mass $m$ and charge $q$ 
in the linear approximation is the same as the inertial
force experienced by the free particle with respect to an accelerated 
system of reference with translational acceleration
${\bf a}_L=-q_E{\bf E}/m$ and frequency of rotation 
$\boldsymbol{\Omega}_L=q_B{\bf B}/(2mc)$. In electromagnetism 
$q_E=q_B=q$
and for all particles with the same charge-to-mass ratio $q/m$, the 
electromagnetic field can be replaced by the same
accelerated system. This circumstance takes on a universal character 
in the case of gravity since the gravitational
charge-to-mass ratio is the same for all particles according to the 
principle of equivalence of gravitational and inertial
masses. The universality of the gravitational interaction thus leads 
to a geometric theory of gravitation, i.e. general
relativity. An analogous approach to electrodynamics is impossible 
due to the fact that $q/m$ for different particles can, for
instance, be positive, negative or zero.

To develop a gravitational analogue of Larmor's theorem, we recall 
from the previous section that in the linear approximation
of general relativity, the exterior gravity of a rotating source can 
be described in terms of GEM fields. In a sufficiently
small neighborhood of the exterior region, the GEM fields may be 
considered locally uniform. These fields may then be locally
replaced by an accelerated system in Minkowski spacetime. To this 
end, let us imagine an accelerated observer following a
worldline $x^\mu_0(\tau )$. Here
$\tau
$ is the observer's proper time and
$u^\mu=dx_0^\mu /d\tau
$ and $A^\mu=du^\mu /d\tau$ are its velocity and acceleration 
vectors, respectively. Let $\lambda^\mu _{\;\; (\alpha)}$ be the
orthonormal tetrad frame of the observer such that 
$\lambda^\mu_{\;\;(0)}=u^\mu$ and
\begin{equation}\label{eq19} 
\frac{d\lambda^\mu_{\;\;(\alpha)}}{d\tau} =\phi_\alpha ^{\;\; \beta} 
\lambda^\mu_{\;\; (\beta
)},\end{equation}
where $\phi_{\alpha \beta}(\tau )$ is the antisymmetric acceleration 
tensor of the observer. In analogy with the Faraday
tensor, $\phi_{\alpha\beta}$ consists of an``electric" part 
$\phi_{0i}=a_i$ and a ``magnetic" part
$\phi_{ij}=\epsilon_{ijk}\Omega^k$. Here ${\bf a}$ and 
$\boldsymbol{\Omega}$ are spacetime scalars that represent 
respectively
the translational acceleration, $a_i=a_\mu \lambda^\mu_{\;\; (i)}$, 
and the rotational frequency of the local spatial frame
with respect to the local nonrotating (i.e. Fermi-Walker transported) 
frame. Consider now a geodesic system of coordinates
$X^\mu$ established along the worldline of the fiducial observer. At 
any event $\tau$ along the worldline, the straight
spacelike geodesic lines orthogonal to the worldline span a 
hyperplane that is Euclidean space. Let $x^\mu$ be the
coordinates of a point on this hyperplane; then,
\begin{equation}\label{eq20} x^\mu =x^\mu_0+X^i\lambda^\mu_{\;\; 
(i)}(\tau ),\quad \tau =X^0.\end{equation}
The Minkowski metric $\eta _{\mu\nu}dx^\mu dx^\nu$ with respect to 
the new coordinates takes the form
$g_{\mu\nu}dX^\mu dX^\nu$, where
\begin{align}\label{eq21} g_{00}&=-(1+{\bf a}\cdot {\bf 
X})^2+(\boldsymbol{\Omega}\times {\bf X})^2,\\
\label{eq22} g_{0i}&=(\boldsymbol{\Omega}\times {\bf X})_i,\quad 
g_{ij}=\delta_{ij}.\end{align}
These geodesic coordinates are admissible if $g_{00}<0$; a detailed 
discussion of the nature of the boundary of the admissible
region is given in \cite{12}.

A comparison of the metric given by \eqref{eq21} and \eqref{eq22} 
with \eqref{eq4} reveals that they are Larmor equivalent at
the linear order once
\begin{equation}\label{eq23} \Phi=-{\bf a}_L\cdot {\bf X},\quad {\bf 
A}=-\frac{1}{2}\boldsymbol{\Omega}_{L} \times {\bf
X},\end{equation}
and we neglect spatial curvature. The corresponding GEM fields to 
lowest order are ${\bf E}=-\boldsymbol{\nabla} \Phi ={\bf
a}_L$ and
${\bf B}=\boldsymbol{\nabla}\times {\bf A}=-\boldsymbol{\Omega}_L$, 
as expected from the traditional Larmor theorem with
$q_E=-m$ and $q_B=-2m$.

The gravitational Larmor theorem \cite{13} is essentially Einstein's 
principle of equivalence formulated within the GEM
framework. Einstein's heuristic principle of equivalence 
traditionally refers to the Einstein ``elevator'' and its
translational acceleration in connection with the gravitoelectric 
field of the source. However, it follows from the
gravitational Larmor theorem that a rotation of the elevator is 
generally necessary as well in order to take due account of
the gravitomagnetic field of the source.

In classical electrodynamics, a charged spinning test particle has a 
magnetic dipole moment $\boldsymbol{\mu}=q{\bf S}/(2mc)$,
where $m$, $q$ and ${\bf S}$ are respectively the mass, charge and 
the spin of the particle. In an external magnetic field
${\bf B}$, the test dipole has an interaction energy 
$-\boldsymbol{\mu }\cdot{\bf B}$ and precesses due to a torque
$\boldsymbol{\mu}\times {\bf B}$. In a similar way, a test gyroscope 
of spin ${\bf S}$ with $q\to q_B=-2m$ has a
gravitomagnetic dipole moment $\boldsymbol{\mu}_g=-{\bf S}/c$ and 
precesses in the exterior field of a rotating source of mass
$M$ and spin ${\bf J}$ with the frequency \cite{14}
\begin{equation}\label{eq24}\boldsymbol{\Omega}_P=\frac{GJ}{c^2r^3}[3(\hat{{\bf 
J}}\cdot \hat{{\bf r}})\hat{{\bf r}}-\hat{{\bf
J}} ],\end{equation}
where $\boldsymbol{\Omega}_P={\bf B}/c$ and the gravitomagnetic field 
${\bf B}$ is given by the curl of the vector potential
in \eqref{eq5} that corresponds to a source with gravitomagnetic 
dipole moment ${\bf J}/c$. The related interaction energy is
${\bf S}\cdot \boldsymbol{\Omega}_P$. A major aim of the GP-B is to 
measure \eqref{eq24} for gyros in a polar orbit about the
Earth.

It follows from \eqref{eq24} that $2\pi /\Omega_P$ is a 
characteristic timescale for the gravitomagnetic field. More
generally, gravitomagnetic effects reveal an interesting temporal 
structure around a rotating mass; this can be further
illustrated by the phenomena associated with the gravitomagnetic 
clock effect \cite{15} and the gravitomagnetic time delay
\cite{16}.

A more exact long-term post-Schwarzschild analysis of the orbital 
motion of an ideal test gyroscope in the field of a rotating
source reveals that besides the gravitoelectric geodetic (i.e. de 
Sitter-Fokker) precession of the gyro axis there is a complex
gravitomagnetic component involving precessional as well as 
nutational motions---the latter is known as relativistic nutation
\cite{17}. The net gravitomagnetic spin motion reduces in the 
post-Newtonian approximation to equation \eqref{eq24}.

\section{Spacetime Curvature Approach to GEM}

The main elements that underlie the gravitational Larmor theorem 
apply equally well but in a different context to an
alternative treatment of GEM based on spacetime curvature. Unlike the 
previous treatment, the new approach is not limited to
perturbations of flat spacetime and can be employed in an arbitrary 
curved spacetime.

Consider a congruence of test observers following geodesics in a 
gravitational field. Choosing a reference observer in this
congruence, we set up a Fermi coordinate system along its path. This 
amounts to constructing an inertial system of coordinates
in the immediate neighborhood of the reference observer \cite{18}. 
Let $\lambda^\mu_{\;\; (\alpha)}(\tau )$ be the
orthonormal tetrad of the reference observer. Here $\lambda^\mu_{\;\; 
(0)}$ is the vector tangent to the worldline of the
observer and $\tau$ is the proper time along its path. The spatial 
frame $\lambda^\mu_{\;\; (i)}$, $i=1,2,3$, consists of
unit vectors along ideal gyro directions that are parallel 
transported along the worldline. The Fermi frame is a geodesic
reference system that is based on the nonrotating orthonormal tetrad 
$\lambda^\mu_{\;\; (\alpha )}$. The metric of spacetime
in Fermi coordinates $X^\mu =(T,{\bf X})$ is then given by
\begin{align}\label{eq25} g_{00}&=-1-R_{0i0j}X^iX^j+\ldots ,\\
\label{eq26} g_{0i}&=-\frac{2}{3}R_{0jik}X^jX^k+\ldots ,\\
\label{eq27} g_{ij}&=\delta_{ij}-\frac{1}{3}R_{ikjl}X^kX^l+\ldots ,\end{align}
where $R_{\alpha \beta \gamma \delta}(T)$ is the projection of the 
Riemann curvature tensor on the orthonormal tetrad of the
reference observer
\begin{equation}\label{eq28} R_{\alpha \beta \gamma \delta}=R_{\mu\nu \rho
\sigma}\lambda^\mu_{(\alpha)}\lambda^\nu_{(\beta)}\lambda^\rho_{(\gamma)}\lambda^\sigma 
_{(\delta)}.\end{equation}
Along the reference geodesic $T=\tau$, ${\bf X}=0$ and 
$g_{\mu\nu}=\eta_{\mu\nu}$ by construction. The Fermi coordinates are
admissible within a cylindrical spacetime region of radius $\sim 
\mathcal{R}$ around the worldline of the reference observer.
Here $\mathcal{R}$ is the radius of curvature of spacetime.

The metric in Fermi coordinates---within the limited region of 
admissibility---has the form of a perturbation about Minkowski
spacetime; therefore, using the previous GEM approach and comparing 
\eqref{eq25} and \eqref{eq26} with \eqref{eq4}, we define
the new GEM potentials as
\begin{align}\label{eq29} \Phi (T, {\bf X})&=-\frac{1}{2}R_{0i0j}(T 
)X^iX^j+\ldots ,\\
\label{eq30} A_i(T ,{\bf X})&=\frac{1}{3}R_{0jik}(T )X^jX^k+\ldots ,\end{align}
where the spatial curvature has been ignored. The GEM fields are 
defined in terms of the potentials as before and are given to
lowest order as
\begin{align}\label{eq31} E_i(T ,{\bf X})&=R_{0i0j}(T) X^j+\ldots ,\\
\label{eq32} B_i(T ,{\bf 
X})&=-\frac{1}{2}\epsilon_{ijk}R^{jk}_{\;\;\;\; 0l}(T )X^l +\ldots 
.\end{align}

It is interesting to note that in this approach the gravitoelectric 
field is directly connected with the ``electric''
components of the curvature tensor $R_{0i0j}$ and the gravitomagnetic 
field is directly connected with the ``magnetic''
components of the curvature tensor $R_{0ijk} $ \cite{19}. It is 
possible to combine \eqref{eq31} and \eqref{eq32} in the GEM
Faraday tensor
\begin{equation}\label{eq33} F_{\alpha \beta}=-R_{\alpha \beta 
0i}X^i\end{equation}
to linear order in ${\bf X}$, where $F_{0i}=-E_i$ and 
$F_{ij}=\epsilon _{ijk}B^k$. Then, Maxwell's equations $F_{[\alpha 
\beta
,\gamma]}=0$ and $F^{\alpha \beta}_{\;\;\;\; ,\beta}=4\pi J^\alpha$ 
are satisfied in this case to lowest order in $|{\bf
X}|/\mathcal{R}$ with
\begin{equation}\label{eq34} 4\pi J_\alpha (T ,\mathbf{0})=-R_{0\alpha}=-8\pi 
G\left(T_{0\alpha
}-\frac{1}{2}\eta_{0\alpha}T^\beta_{\;\; \beta}\right)\end{equation}
along the reference trajectory in Fermi coordinates. Here we have 
used the gravitational field equations as well as the
symmetries of the Riemann curvature tensor.

The new approach to GEM naturally contains the analogue of the 
Lorentz force law. The motion of free test particles in the
congruence relative to the reference particle at the spatial origin 
of Fermi coordinates can be expressed as
\begin{equation}\begin{split}\label{eq35} 
\frac{d^2X^i}{dT^2}&+R_{0i0j}X^j+2R_{ikj0}V^kX^j+\left(2R_{0kj0}V^iV^k\right.\\
&+\left.\frac{2}{3}R_{ikjl}V^kV^l+\frac{2}{3} 
R_{0kjl}V^iV^kV^l\right) X^j=0,\end{split}\end{equation}
valid to linear order in the separation ${\bf X}$. This geodesic 
deviation equation is a generalized Jacobi equation
\cite{20} in which the rate of geodesic separation (i.e. the relative 
velocity of the test particle) ${\bf V}=d{\bf X}/dT$ is
in general arbitrary ($|{\bf V}|<1$ at $\mathbf{X}=\mathbf{0}$). To linear order in velocity, 
one can show that \eqref{eq35} takes the Lorentz form
\begin{equation}\label{eq36} m\frac{d^2{\bf X}}{dT^2}=q_E{\bf 
E}+q_B{\bf V}\times {\bf B},\end{equation}
where $q_E=-m$ and $q_B=-2m$ as before.

The stress-energy tensor in the new approach can be constructed 
essentially from the Faraday tensor \eqref{eq33} as in
Maxwell's theory, i.e.
\begin{equation}\label{eq37}G\mathcal{T}^{\alpha 
\beta}=\frac{1}{4\pi}\left(F^\alpha _{\;\; \gamma }F^{\beta
\gamma}-\frac{1}{4} g^{\alpha \beta} F_{\gamma \delta }F^{\gamma 
\delta}\right),\end{equation}
where an extra factor of $G$ has been introduced due to dimensional 
considerations. From equations \eqref{eq33} and
\eqref{eq37}, we find
\begin{equation}\label{eq38} \mathcal{T}^{\alpha \beta}=\frac{1}{4\pi 
G}\left( R^\alpha_{\;\; \gamma 0i}R^{\beta \gamma
}_{\;\;\;\; 0j}-\frac{1}{4}\eta^{\alpha \beta}R_{\gamma \delta 
0i}R^{\gamma \delta}_{\;\;\;\; 0j}\right)X^iX^j.\end{equation}
This tensor as well as the Faraday tensor \eqref{eq33} vanishes along 
the worldline of the reference observer in the Fermi
system; indeed, this is an immediate consequence of the inertial 
character of this system along the reference trajectory and a
realization of Einstein's principle of equivalence. Thus this 
treatment depends on our choice of a reference observer and the
corresponding Fermi coordinate system.

To obtain a coordinate-independent measure of the stress-energy 
content of the gravitational field, we invoke the notion that
the physical measurement of such a quantity requires an averaging 
process \cite{21}. Starting from an event $(T,{\bf 0})$ on
the reference worldline, we average the tensor given in \eqref{eq38} 
over a small sphere of radius $\epsilon L$, where
$0<\epsilon <<1$ and $L$ is an invariant length scale that is 
characteristic of the source of the gravitational field under
consideration. For instance, $L$ could be $GM/c^2$ or, in the absence 
of such a scale, the Planck length. The quadratic nature
of \eqref{eq38} in the spatial coordinates implies that the averaging involves
\begin{equation}\label{eq39} \langle X^iX^j\rangle =k(\epsilon 
L)^2\delta_{ij},\end{equation} where $k=1/5$ or $1/3$ depending
on whether the averaging involves the volume or the surface of the 
sphere, respectively. In either case, the constant $k$ can
be absorbed in the definition of $L$. Thus
\begin{equation}\label{eq40} \langle \mathcal{T}_{\alpha \beta 
}\rangle =\frac{k\epsilon^2L^2}{4\pi G}\bar{T}_{\mu\nu \rho
\sigma} \lambda 
^\mu_{(\alpha)}\lambda^\nu_{(\beta)}\lambda^\rho_{(0)} \lambda^\sigma 
_{(0)},\end{equation}
where $\bar{T}_{\mu\nu \rho \sigma }(x)$ is the Bel tensor given by
\begin{equation}\label{eq41}\bar{T}_{\mu\nu\rho 
\sigma}(x)=\frac{1}{2}(R_{\mu\xi\rho \zeta}R_{\nu \;\; \rho}^{\;\; 
\xi \;\;
\zeta }+R_{\mu \xi \sigma \zeta} R_{\nu \;\; \rho}^{\;\; \xi \;\; 
\zeta})-\frac{1}{4} g_{\mu\nu}R_{\alpha \beta \rho \gamma
}R^{\alpha
\beta
\;\;
\gamma}_{\;\;\;\;
\sigma}.\end{equation} This tensor bears a certain similarity with 
the Maxwell stress-energy tensor and on this basis was
first defined by Bel
\cite{22} for Einstein spaces in 1958. The Bel superenergy tensor is 
symmetric and trace-free in its first pair of indices and
only symmetric in the second pair of indices. In a Ricci-flat 
spacetime, the Riemann tensor reduces to the Weyl conformal
tensor
$C_{\mu\nu\rho \sigma}$ and the Bel tensor reduces to the completely 
symmetric and trace-free Bel-Robinson tensor
$T_{\mu\nu\rho \sigma}$ given by
\begin{equation}\label{eq42} T_{\mu\nu\rho \sigma }=\frac{1}{2} 
(C_{\mu\xi\rho \zeta} C_{\nu\;\; \sigma}^{\;\; \xi \;\;
\zeta}+C_{\mu\xi\sigma \zeta}C_{\nu\;\;\rho}^{\;\;\xi \;\; 
\zeta})-\frac{1}{16}g_{\mu\nu} g_{\rho \sigma}C_{\alpha \beta
\gamma \delta}C^{\alpha \beta\gamma \delta}.\end{equation}

An invariant average GEM stress-energy tensor of the gravitational 
field can thus be defined up to a constant positive
multiplicative factor by \cite{23}
\begin{equation}\label{eq43} \bar{T}_{(\alpha 
)(\beta)}=\frac{L^2}{G}\bar{T}_{\mu\nu\rho \sigma 
}\lambda^\mu_{(\alpha)}
\lambda^\nu_{(\beta)}\lambda ^\rho _{(0)} \lambda^\sigma _{(0)},\end{equation}
where $\bar{T}_{(\alpha )(\beta)}$ is symmetric and traceless. In the 
Ricci-flat case, $\bar{T}_{(\alpha )(\beta)}\to
T^*_{(\alpha )(\beta)}$, where $T^*_{(\alpha )(\beta)}$ is the {\it 
gravitational stress-energy tensor}. This designation
refers to the fact that in a Ricci-flat spacetime, the spatial 
components of the curvature tensor in \eqref{eq27}, which were
essentially ignored in our GEM analysis, are indeed basically given 
by its electric components; therefore, $T^*_{(\alpha
)(\beta)}$ involves all of the components of the spacetime curvature. 
The stress-energy tensors $\bar{T}_{\mu\nu}$ and
$T^*_{\mu\nu}$ have properties reminiscent of Maxwell's 
electrodynamics and have been discussed in detail in \cite{23}. For
instance, in the exterior of a rotating mass the gravitational 
Poynting flux based on the new approach has a flow velocity
given by \eqref{eq18} with $k=3$.

A significant generalization of the concept of superenergy tensors 
has been developed in \cite{24}.

The gravitomagnetic contribution to the spacetime curvature due to a 
rotating source involves subtle cumulative effects
\cite{25} that can be measured in principle via relativistic gravity 
gradiometry \cite{26}.

\section{Spin-Rotation-Gravity Coupling}

An issue of fundamental interest is whether intrinsic spin is 
affected by a gravitomagnetic field in basically the same way as
the classical spin of an ideal gyroscope. This question is related to 
the inertia of intrinsic spin. The description of
physical states in the quantum theory is based upon the irreducible 
unitary representations of the inhomogeneous Lorentz
group, which are characterized by means of mass and spin. The 
inertial properties of mass in moving frames of reference are
already well known: for instance, via Coriolis, centrifugal and other 
mechanical effects, as well as their quantum mechanical
counterparts. The inertial properties of intrinsic spin involve the 
phenomena associated with the spin-rotation-gravity
coupling.

The coupling of intrinsic spin with rotation reveals the rotational 
inertia of intrinsic spin. That is, as a particle moves,
its intrinsic spin keeps its aspect with respect to an inertial 
frame; therefore, the spin appears to rotate with respect to a
rotating observer. To this motion of spin corresponds, according to 
quantum mechanics, a Hamiltonian $H=-\gamma
\boldsymbol{\Omega }\cdot {\bf S}$. The general formula for the 
transformation of energy turns out to be $\mathcal{E}'=\gamma
(\mathcal{E}-\hbar \Omega \mathcal{M})$, where $\Omega$ is the frequency of 
rotation of the observer and $\hbar \mathcal{M}$ is the
component of the total angular momentum along the axis of rotation; 
that is, $\mathcal{M}=0,\pm 1,\pm 2,\ldots$ for a scalar or
a vector particle, while $\mathcal{M}\mp \frac{1}{2}=0, \pm 1,\pm 
2,\ldots $ for a Dirac particle. This formula relates the
energy of a quantum system measured by a rotating observer 
$\mathcal{E}'$ to measurements performed in a global inertial frame
and can be written in the JWKB approximation as $\mathcal{E}'=\gamma 
(\mathcal{E}-\boldsymbol{\Omega}\cdot {\bf J})$, where
${\bf J}= {\bf r}\times {\bf p}+{\bf S}$ is the total angular 
momentum. Thus, $\mathcal{E}'=\gamma (\mathcal{E}-{\bf v}\cdot
{\bf p})-\gamma \boldsymbol{\Omega }\cdot {\bf S}$, so that in the 
absence of intrinsic spin we recover the classical
expression for the energy of a particle as measured in the rotating 
frame with ${\bf v}=\boldsymbol{\Omega}\times {\bf r}$.
The spin-rotation coupling therefore involves an energy shift given 
by the Hamiltonian $H=-\gamma \boldsymbol{\Omega
}\cdot {\bf S}$
\cite{27}.

Observational evidence for such an energy shift in the case of 
fermions has been provided in certain high-precision
experiments by way of a small frequency offset due to the coupling 
between the nuclear spin of mercury and the rotation of the
Earth \cite{28,29}; moreover, a direct approach using neutron or atom 
interferometry has been proposed in \cite{30}. For
photons, helicity-rotation coupling has been confirmed to rather high 
accuracy using rotating GPS receivers \cite{31};
moreover, experimental evidence exists for such a coupling in the 
microwave and optical regimes in terms of the frequency
shift of polarized radiation \cite{30}. The modifications of Doppler 
effect and aberration due to the coupling of photon spin
with the rotation of the source and/or receiver have been the subject 
of recent studies \cite{32,33}.

Let us now turn to the coupling of spin with gravitomagnetic fields; 
the spin-gravity coupling is naturally related to the
spin-rotation coupling by way of Einstein's principle of equivalence. 
That is, starting from the spin-rotation Hamiltonian,
the transformation $\boldsymbol{\Omega }\to -\boldsymbol{\Omega}_P$ 
leads, according to the gravitational Larmor theorem, to
the spin-gravity Hamiltonian.

It follows from  these ideas that in Earth-based experiments, to 
every Hamiltonian we must add the spin-rotation-gravity
interaction Hamiltonian $\delta H \cong -\boldsymbol{\Omega} _\oplus 
\cdot {\bf S}+ \boldsymbol{\Omega}_P\cdot{\bf S}$,
where
$\boldsymbol{\Omega }_\oplus$ and $\boldsymbol{\Omega}_P$ refer to 
the rotation frequency of the Earth and the corresponding
gravitomagnetic precession frequency, respectively. Thus in the 
approximation under consideration here a particle with
intrinsic spin behaves essentially like an ideal gyroscope. The 
energy difference corresponding to a spin-$1/2$ particle
polarized vertically up and down relative to the surface of the Earth 
can be estimated from $\hbar \Omega_\oplus \cong
10^{-19}\text{eV}$ and $\hbar \Omega_P\cong 10^{-29}\text{eV}$. The 
measurement of the latter term is beyond present
capabilities by several orders of magnitude. In this connection, 
however, we note that near Jupiter $\hbar \Omega_P\cong
10^{-27}\text{eV}$, and therefore it is likely that with further 
improvements in magnetometer design, the spin-gravitomagnetic
coupling could become measurable in a satellite in orbit near the 
surface of Jupiter in the foreseeable future \cite{34}. It
is important to recognize that such a relativistic quantum 
gravitational effect, like all other gravitational effects, is
subject to the whole mass-energy content of the universe. It follows 
that our treatment has been based on certain
cosmological assumptions regarding the distribution of angular 
momentum in the universe; specifically, we have assumed that on
the largest scales there is no preferred sense of rotation. Moreover, 
in $\delta H$ the spin-gravity coupling term has a
gradient. Therefore, there exists a gravitomagnetic Stern-Gerlach 
force $-\nabla (\boldsymbol{\Omega}_P\cdot {\bf S})$ on a
spinning particle that is independent of the its mass and hence 
violates the universality of the gravitational acceleration.
The weight of a body thus depends on its spin, but the effect is too 
small to be directly measurable in the foreseeable
future. It is interesting to note that the Stern-Gerlach force has an 
exact analogue in the classical Mathisson-Papapetrou
spin-curvature force. The results of this section are in agreement 
with the consequences of Dirac-type wave equations in the
gravitational field of a rotating mass
\cite{35}.

\end{document}